\documentclass[useAMS,usenatbib,usegraphicx]{mn2e}
\usepackage{amsmath}
\usepackage{color}
\usepackage{hyperref}

\newif\ifAMStwofonts
\AMStwofontstrue

\newcommand{\simlt}{\lower.5ex\hbox{$\; \buildrel < \over \sim \;$}}
\newcommand{\simgt}{\lower.5ex\hbox{$\; \buildrel > \over \sim \;$}}
\newcommand{\kms}{km\,s$^{-1}$}

\newcommand{\apj}{ApJ}

\newcommand{\apjs}{ApJS}

\newcommand{\mnras}{MNRAS}

\newcommand{\aj}{AJ}

\title[Exploring a new definition of the green valley and its implications]
{Exploring a new definition of the green valley and its implications}
\author[Angthopo, Ferreras \& Silk]
{James Angthopo$^{1}$, Ignacio Ferreras$^{2,3,4}$\thanks{E-mail: i.ferreras@ucl.ac.uk}, Joseph Silk$^{5,6,7}$
\\
$^1$ Mullard Space Science Laboratory, University College London, 
Holmbury St Mary, Dorking, Surrey RH5 6NT, UK\\
$^2$ Department of Physics and Astronomy, University College London,
Gower Street, London WC1E 6BT, UK\\
$^3$ Instituto de Astrof{\'i}sica de Canarias, Calle V{\'i}a L{\'a}ctea s/n,
E38205, La Laguna, Tenerife, Spain\\
$^4$ Departamento de Astrof{\'i}sica, Universidad de La Laguna (ULL), E-38206 La Laguna,
Tenerife, Spain\\
$^5$ Institut d'Astrophysique de Paris (UMR 7095: CNRS \& UPMC), 98 bis Bd Arago, F-75014 Paris, France\\
$^6$ Sub-department of Astrophysics, University of Oxford, Keble Road, Oxford OX1 3RH, UK\\
$^7$ Department of Physics and Astronomy, The Johns Hopkins University Homewood Campus, Baltimore, MD 21218, USA
}

\begin{document}
\date{MNRAS Letters. Accepted 2019 June 29. Received 2019 June 28; in original form 2019 February 25}
\pagerange{\pageref{firstpage}--\pageref{lastpage}} \pubyear{2019}
\maketitle
\label{firstpage}

%%\newif\ifAMStwofonts
%%\AMStwofontstrue

\begin{abstract}
The distribution of galaxies on a colour-magnitude diagram reveals a
bimodality, featuring a passively evolving red sequence and a star-forming
blue cloud. The region between these two,
the Green Valley (GV), represents a fundamental transition
where quenching processes operate.  We exploit an alternative
definition of the GV using the 4,000\,\AA\ break strength, an indicator that is 
more resilient than colour to dust attenuation. 
We compare and contrast our GV definition with the traditional one, based on
dust-corrected colour, making  use of data from the Sloan Digital
Sky Survey. Our GV selection -- that does not need a dust correction and thus does not
carry the inherent systematics -- reveals very similar trends regarding
nebular activity (star formation, AGN, quiescence) to the standard 
dust-corrected $^{0.1}(g-r)$. By use of high SNR stacked spectra of the quiescent
GV subsample, we derive the simple stellar population (SSP) age
difference across the GV, a rough proxy of the quenching timescale
($\Delta$t).  We obtain an increasing trend with velocity dispersion
($\sigma$), from $\Delta$t$\sim$1.5\,Gyr at $\sigma$=100\,\kms, up to
3.5\,Gyr at $\sigma$=200\,\kms, followed by a rapid decrease in the
most massive GV galaxies ($\Delta$t$\sim$1\,Gyr at
$\sigma$=250\,\kms), suggesting two different modes of quenching,
or the presence of an additional channel (rejuvenation).
\end{abstract} 

<

%%%%%%%%%%%%%%%%%%%%%%%%%%%%%%%%%%%%%%%%%%%%%%%%
\section{Introduction}
\label{Sec:Intro}

Broadband photometry is widely used to assess the properties of the
underlying stellar populations of galaxies. Among the many diagnostic
diagrams adopted to study galaxy evolution are the colour-magnitude, 
or colour-mass, diagram \citep[e.g.,][hereafter M07]{Martin:07}, and the UVJ bi-colour
plots \citep[e.g.,][]{Whitaker:11}. These diagrams allow us to separate
galaxies with respect to the stellar age distribution. However, a colour-based selection 
requires correcting for dust attenuation, possibly introducing 
a model-dependent systematic, especially as the attenuation law is found
to vary widely among galaxies (e.g., \citealt{KC:13}, \citealt{Nara:18}; \citealt{TMF:18}).
An alternative approach would ideally involve the use of age-sensitive
indicators that are not so severely affected by dust, suggesting the use of
spectral features that could be measured on relativelly low S/N spectra.

The distribution of galaxies on a colour vs magnitude (or vs stellar
mass) diagram reveals a conspicuous
bimodality \citep{Strateva:01,Baldry:04,Bell:04,Mateus:06}, featuring a 
Red Sequence (RS), mostly dominated by old and passive galaxies, and a
Blue Cloud (BC), made of star-forming galaxies. The area separating these two
populations, termed the Green Valley (GV), represents a
transition region where star formation is
quenched \citep{Menci:05,Faber:07,Schawinski:07,Goncalves:12,Salim:14}.  However, the
interpretation of GV galaxies and the process of quenching
remains one of the key open questions of galaxy
evolution \citep{Schawinski:14,Ned:15,Bremer:18,Eales:18}.

%%%%%%%%%%%%%%%%%%%%%%%%%%%%%%%%%%%%%%%%%%%%%%%%%%%%%%%%%%%%%%%%%%%%%
%%%%%%%%%%%   FIGURE 1
%%%%%%%%%%%%%%%%%%%%%%%%%%%%%%%%%%%%%%%%%%%%%%%%%%%%%%%%%%%%%%%%%%%%%
\begin{figure*}
  \centering
  \includegraphics[width=80mm]{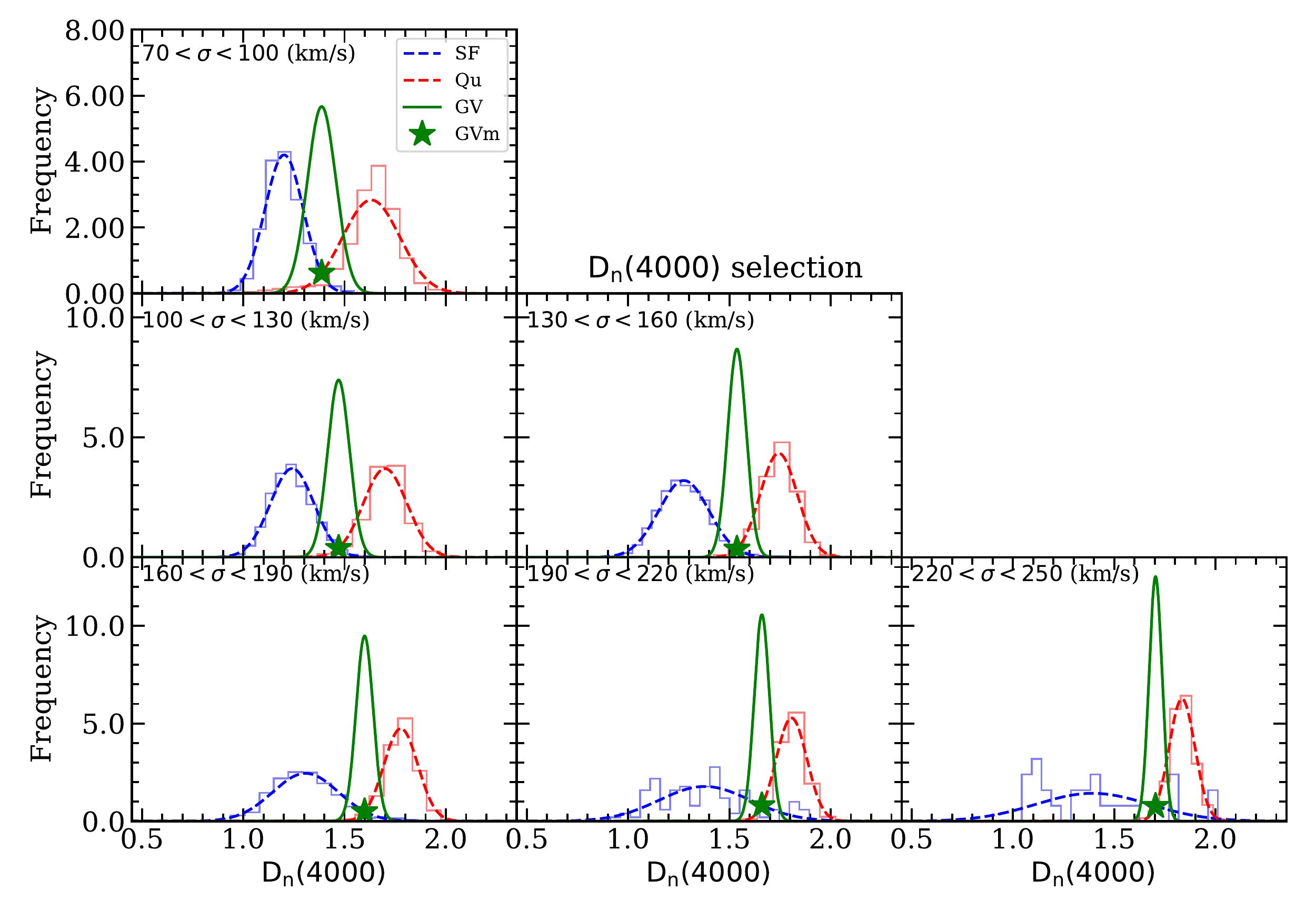}
  \includegraphics[width=80mm]{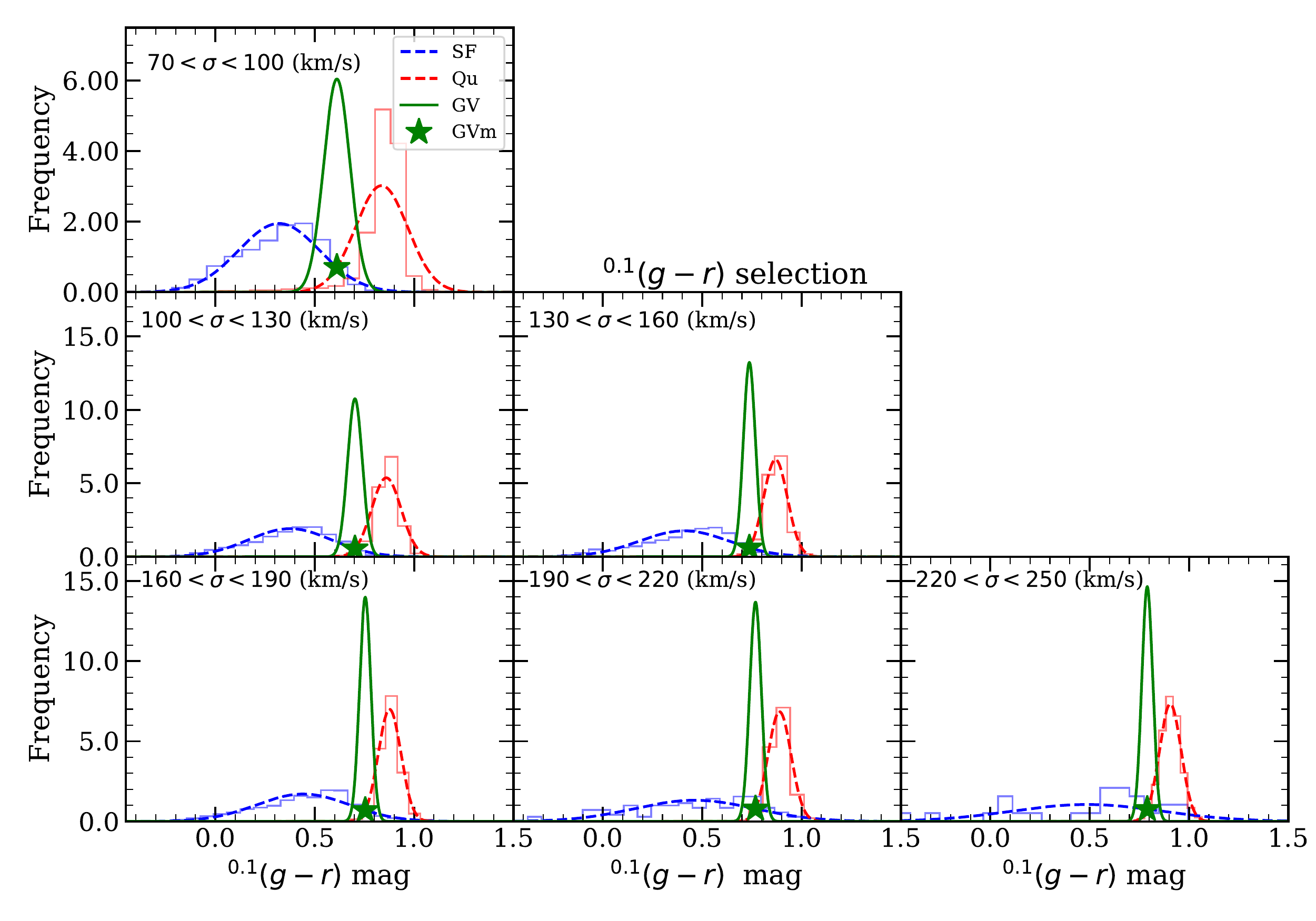}
  \caption{Definition of the GV population: The distribution
    of star-forming (blue) and quiescent galaxies (red) is used to
    define a probability distribution function (PDF) for the BC
    (${\cal P}_{\rm BC}$) and RS (${\cal P}_{\rm  RS}$)
    , respectively. A Gaussian approximation is adopted. The
    place where ${\cal P}_{\rm BC}={\cal P}_{\rm RS}$ defines the
    mean point of the GV, shown by a star symbol. The green line maps
    the expected PDF of GV galaxies. The left  panels
    correspond to a selection based on 4,000\AA\ break
    strength, and the right panels adopt $^{0.1}(g-r)$ colour
    (dust-corrected) to define the GV.}
  \label{fig:GV_PDF}
\end{figure*}
%%%%%%%%%%%%%%%%%%%%%%%%%%%%%%%%%%%%%%%%%%%%%%%%%%%%%%%%%%%%%%%%%%%%%

%%%%%%%%%%%%%%%%%%%%%%%%%%%%%%%%%%%%%%%%%%%%%%%%%%%%%%%%%%%%%%%%%%%%%
%%%%%%%%%%%   FIGURE 2
%%%%%%%%%%%%%%%%%%%%%%%%%%%%%%%%%%%%%%%%%%%%%%%%%%%%%%%%%%%%%%%%%%%%%
\begin{figure*}
  \centering
  \includegraphics[width=155mm]{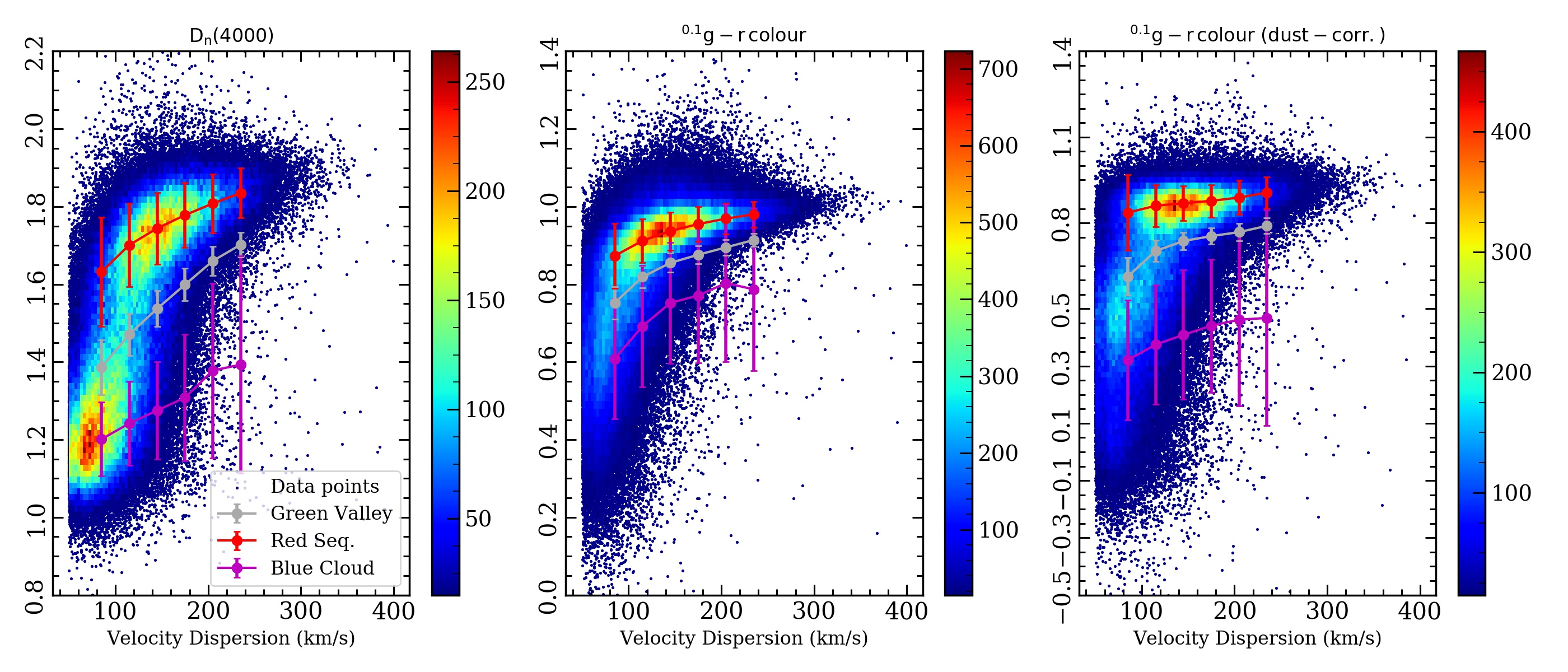}
  \caption{{\it Left:} Distribution of SDSS galaxies with high
    S/N spectra on the D$_n$(4000) vs velocity dispersion plane.  The
    colour map traces number density of individual galaxies on the
    diagram (colour bar to the right). The BC, GV
    and RS subsamples are shown, as labelled, with the
    points locating the mean of the GV distribution and the error bars
    spanning the standard deviation (i.e. the corresponding green
    Gaussians in figure~\ref{fig:GV_PDF}). {\it Centre:} Distribution
    of galaxies on the $^{0.1}(g-r)$ colour (dust-uncorrected, shown for reference)
    vs velocity dispersion plane. 
    {\it Right:} Equivalent
    distribution with the dust-corrected $^{0.1}(g-r)$ colour. Note the D$_n$(4000)
    measurements have not been corrected for dust, as the correction is negligible
    for our purposes.}
  \label{fig:Selec}
\end{figure*}
%%%%%%%%%%%%%%%%%%%%%%%%%%%%%%%%%%%%%%%%%%%%%%%%%%%%%%%%%%%%%%%%%%%%%

%%%%%%%%%%%%%%%%%%%%%%%%%%%%%%%%%%%%%%%%%%%%%%%%
\section{Sample}
\label{Sec:Sample}

We make use of a large set of high quality spectra
from the Sloan Digital Sky Survey (SDSS) to define the GV using the
standard D$_n$(4000) feature \citep{Balogh:99} that quantifies the
4,000\AA\ break strength. This selection is compared with the 
standard (colour-based) definition of the GV.
We select from the SDSS Data Release 14 all individual galaxy
spectra \citep{DR14} with a median S/N in the SDSS $r$ band above 10,
within the redshift interval 0.05$<$z$<$0.10. The original data set 
comprises 228,880 spectra.  We choose
the stellar velocity dispersion as the main stacking parameter,
instead of the standard choice of either the total luminosity or
the stellar mass of the galaxy.  The velocity dispersion
($\sigma$) is found to correlate more strongly with the underlying
population properties \citep[see, e.g.,][]{Bernardi:03,Thomas:05}, and is
less prone to potential systematics, as it is measured in a straightforward way
from the absorption line spectra.
The velocity dispersion can be
mapped onto stellar mass ($M_*$) by use of
$\log M_*/M_\odot = (1.84 \pm 0.03)\log(\sigma/100\,{\rm km\,s}^{-1}) + (10.3 \pm 0.3)$,
with the uncertainties quoted at the 68\% level.  Note also that the
velocity dispersion of the spheroidal component correlates well with
black hole mass, more strongly than does stellar
mass \citep{Gultekin:09}.  This will enable us to refine AGN fractions
in the GV.  The sample is divided into six velocity dispersion bins
between $\sigma$=70 and 250\,km\,s$^{-1}$ in steps of
$\Delta\sigma$=30\,km\,s$^{-1}$.  We note that with the adopted range
of $\sigma$, along with the constraint on the S/N of the spectra, the
estimates of velocity dispersion provided by the SDSS {\sc SpecObj}
catalogue are reliable\footnote{\tt
http://classic.sdss.org/dr7/algorithms/veldisp.html}.

%%%%%%%%%%%%%%%%%%%%%%%%%%%%%%%%%%%%%%%%%%%%%%%%%%%%%%%%%%%%%%%%%%%%%
%%%%%%%%%%%   FIGURE 3
%%%%%%%%%%%%%%%%%%%%%%%%%%%%%%%%%%%%%%%%%%%%%%%%%%%%%%%%%%%%%%%%%%%%%
\begin{figure*}
  \includegraphics[width=150mm]{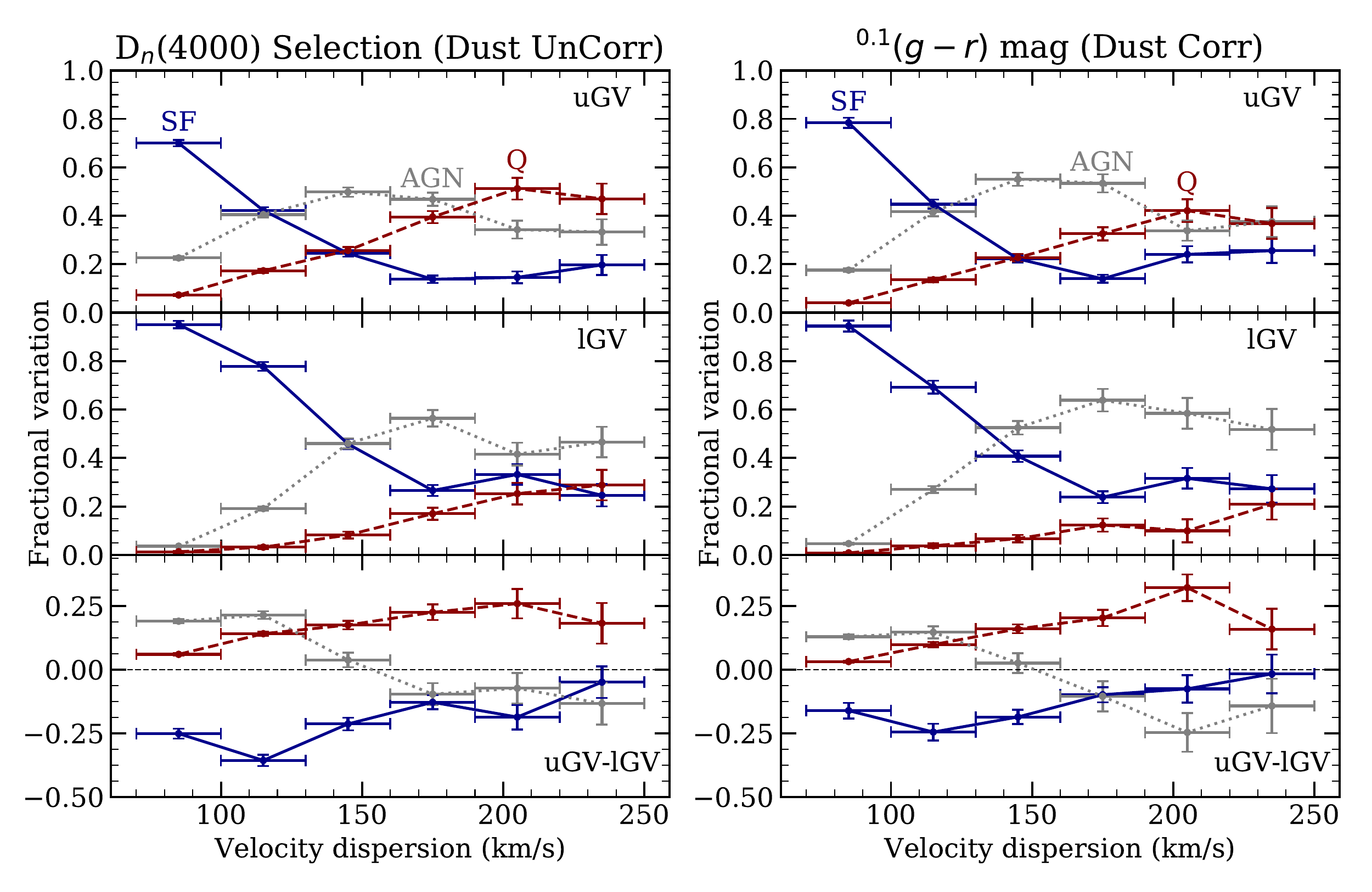}
  \caption{Fraction of galaxy types on the GV. The classification
    is based on the nebular emission properties, as given by the standard
    BPT \citep{BPT:81} diagram, into star-forming (blue);
    AGN/LINER (grey) and quiescent (red). The left (right) panels
    correspond to the selection of GV galaxies based on 4000\AA\ strength  
    (dust-corrected colour). From top to bottom, the panels show the
    fractional contribution according to the spectral type of the
    upper GV, lower GV and the difference upper$-$lower GV.}
  \label{fig:GV_frac}  
\end{figure*}
%%%%%%%%%%%%%%%%%%%%%%%%%%%%%%%%%%%%%%%%%%%%%%%%%%%%%%%%%%%%%%%%%%%%%

%%%%%%%%%%%%%%%%%%%%%%%%%%%%%%%%%%%%%%%%%%%%%%%%
\section{Defining the Green Valley}
\label{Sec:GVal}

We base our definition of the GV on the relative distribution of the
star-forming (SF) and quiescent (Q) subsamples. We make use of the
MPA-JHU SDSS catalogues \citep{Kauff:03,Jarle:04} that provide a
classification of the galaxy spectra from nebular emission lines into
star-forming, AGN, mixed, or quiescent (with the latter class meaning
no emission lines are present). We follow a probability-based approach,
defining a probability distribution function (PDF) for the BC,
${\cal P}_{\rm BC}$, and a PDF for the RS, ${\cal P}_{\rm RS}$. These
functions are assumed to depend on two parameters only, the stellar
velocity dispersion ($\sigma$) and an additional parameter ($\pi$) 
that serves as a proxy of the stellar age of the underlying populations. We will
compare here two choices, the dust-corrected colour -- defined by 
($g-r$) as observed at a fiducial redshift z=0.1 -- and the dust
uncorrected strength of the 4,000\AA\ break. We adopted the dust
correction values from \cite{Kauff:03}, using the \cite{Calz:00}
attenuation law,  thus enabling a fair
comparison to the analysis of M07. Note we use $A_V$, instead of
$A_z$ as baseline for the correction, where we find, at z=0.1,
$A_g^{0.1} = 1.16 A_V^{0.1}$ and $A_r^{0.1} = 0.88 A_V^{0.1}$.
For each choice of the second parameter, $\pi$, we define
the GV as the region where
${\cal P}_{\rm BC}(\sigma,\pi)\sim {\cal P}_{\rm RS}(\sigma,\pi)$, at
fixed velocity dispersion. More specifically, within a velocity
dispersion bin, we use the number of observed spectra of star-forming
and quiescent galaxies to define the PDFs of the BC and RS, respectively,
fitting them to a Gaussian distribution. The
PDF of the GV is then defined as another Gaussian with mean given by
the value of $\pi$ at which 
${\cal P}_{\rm BC}(\sigma,\pi)={\cal P}_{\rm RS}(\sigma,\pi)$, and
standard deviation given by one half of that corresponding to the
PDF of the quiescent subsample.

Figure~\ref{fig:GV_PDF} illustrates the definition of the GV for the
choice of $\pi$=D$_n$(4000) (left panels), or $\pi=^{0.1}(g-r)$
 (dust corrected, right panels), as the population parameter. Both cases show
the well-known trend towards higher break strengths
or redder colours 
as velocity dispersion increases. The mean position of the GV is
represented by a star symbol. Note the significant difference between
the choice of break strength and colour regarding the overlap of the
SF and Q subsets.  A colour-based approach produces more mixed
subsamples that complicates the selection of GV galaxies, whereas the
4,000\AA\ break strength produces a sharper
separation\footnote{Colours such as $(u-g)$ or $(u-r)$ produce very similar
results to $(g-r)$, however we chose the latter as the observations
have markedly lower uncertainties.}. Note, D$_n$(4000) provides an
alternative, clean separation between BC, GV and RS {\sl without requiring
any dust correction}. Therefore, D$_n$(4000) carries a lower 
systematic uncertainty from the modelling associated to these corrections.
This alternative definition of the GV requires
low SNR spectroscopic data, as the 4000 \AA\ break is 
relatively wide and can be measured even at low spectral resolution
\citep[e.g.,][]{Hathi:09,AHC:13}. Our proposed selection of GV galaxies is timely
with future spectroscopic surveys
on the horizon such as WAVES \citep{WAVES:19}, WEAVE \citep{WEAVE:16},
DESI \citep{DESI:16}, MSE \citep{MSE:16} or J-PAS \citep{JPAS:14}. 

Figure~\ref{fig:Selec} shows the location of the three key
evolutionary regions following this methodology, according to
D$_n$(4000) (left panel), $^{0.1}(g-r)$ (uncorrected, centre panel,
shown for reference), or $^{0.1}(g-r)$ (dust-corrected, right
panel). The points with error bars show the mean and standard
deviation of each subsample.  For reference, individual data points
are shown as a density plot.  Note the impact on the GV definition
between the colour distribution before and after the dust correction
is applied.  Comparatively, the equivalent dust correction for
D$_n$(4000) is smaller, resulting in a minor shift of mean in the
definition of the RS, BC and GV subsets, with a maximum
$\Delta$D$_n$(4000)$\sim$\,0.06\,dex. Thus, the 4,000\AA\ break
strength produces a clean representation, even without dust
correction, of the three phases of evolution under study.

Once the GV
is defined, we further separate it into terciles. The lower
tercile represents galaxies closer to the BC, and is
hereafter defined as the lower Green Valley (lGV). The upper tercile
is closer to the RS and is termed the upper Green Valley (uGV). If we
interpret the GV as a transition region where galaxies quench their
star formation, we can portray the lGV and uGV as the starting and
ending stages of this transition, respectively.

%%%%%%%%%%%%%%%%%%%%%%%%%%%%%%%%%%%%%%%%%%%%%%%%
\section{Looking into the Green Valley}
\label{Sec:Discussion}

Figure~\ref{fig:GV_frac} shows the fraction of star-forming, quiescent
and AGN galaxies as a function of velocity dispersion, on the upper
and lower GV, contrasting the difference between a selection based on
break strength (left), or dust-corrected colour (right). Both methods
give the expected decreasing (increasing) trend of star-forming
(quiescent) galaxies towards increasing velocity dispersion,
reflecting the effect of downsizing. At the massive end
($\sigma$=200-250\,km\,s$^{-1}$), the fraction of AGN in the uGV (lGV)
is $\sim$30$\pm$5\% ($\sim$50$\pm$7\%) according to the D$_n$(4000)
definition, comparable to the equivalent fractions from the
dust-corrected colour selection, namely $\sim$37$\pm$6\%
($\sim$51$\pm$8\%). Our estimate of the fractional contribution of AGN
is compatible with the analysis of
M07 \citep[see][Fig. 20]{Martin:07}, further supporting the idea that
the 4,000\AA\ break strength can be accurately used to define the GV
without the need for dust correction.  Moreover, looking at the
distribtuion of the 4,000\AA\ break strength of M07 in their
transition region \citep[see][Fig. 3]{Martin:07}, the index range is
fully compatible with our definition of the GV.

\begin{figure}
  \includegraphics[width=88mm]{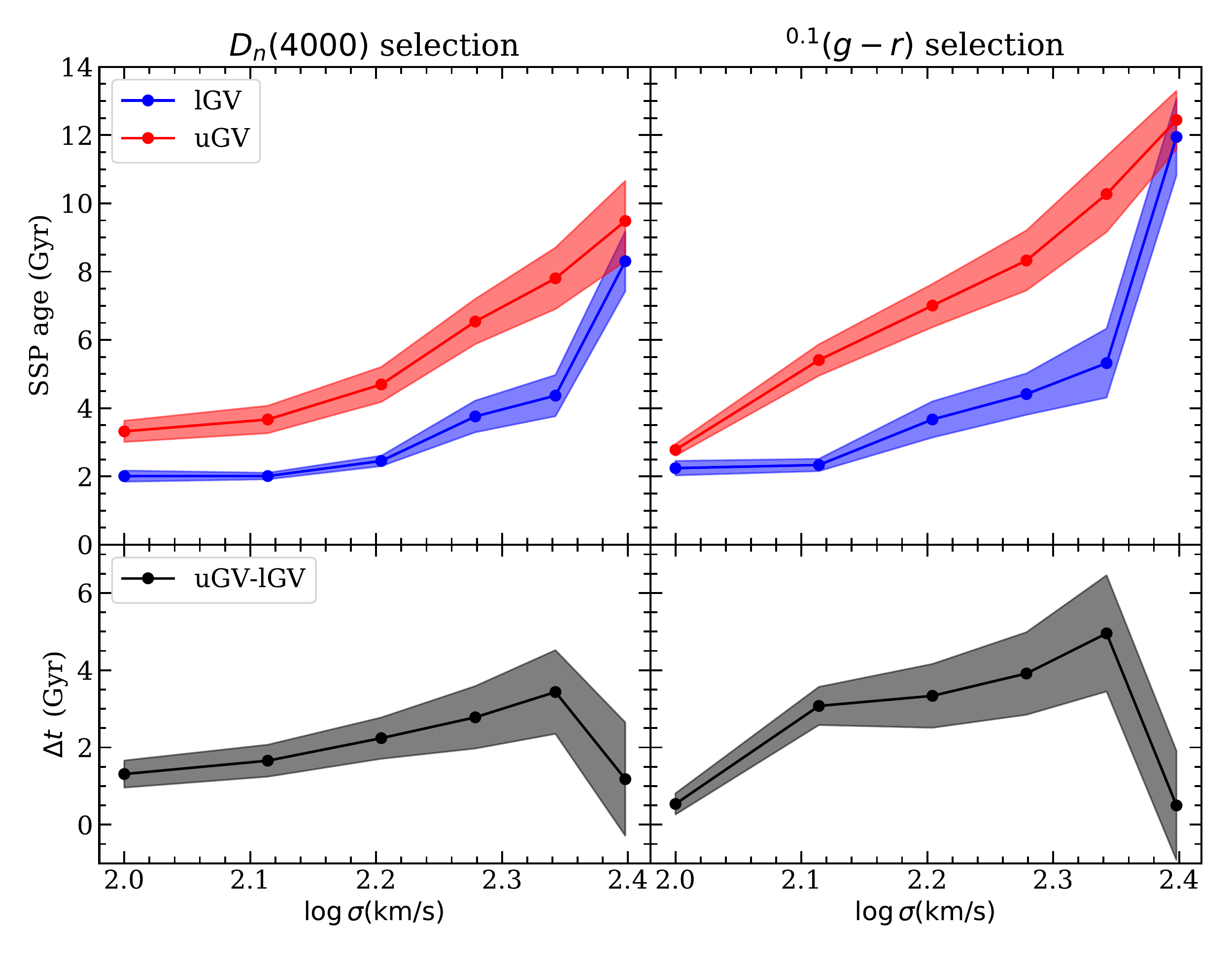}
  \caption{SSP-equivalent age distribution of quiescent galaxies on
    the GV. The left (right) panels correspond to the
    selection of the GV based on 4000\AA\ break strength (dust-corrected colour).
    The top panels show the ages of upper and lower GV galaxies
    separately, whereas the bottom panels give the age difference
    between upper and lower GV at fixed velocity dispersion (defined
    as $\Delta t$). The shaded regions extend over the
    1\,$\sigma$ uncertainties.}
  \label{fig:SSP_age}
\end{figure}

If we view the GV as a transition region where galaxies
evolve from a star-forming state into a passive phase, we can 
compare the stellar age difference, derived from stellar spectra, to
produce an estimate of the timescale expected to traverse this
region. This is a highly non-trivial issue as GV galaxies constitute a
motley distribution of systems, and the presence of star formation
will bias the estimates of stellar age, as all photo-spectroscopic
indicators are inherently luminosity-weighted, thus heavily biased
towards the younger stellar components. The analysis of star-forming
systems typically produce a complex and extended star formation
history.  Note we have avoided AGN galaxies in the definition and
analysis of the GV, as prominent AGN will significantly affect the
spectral continuum of the underlying stellar populations, compromising
the interpretation of the D$_n$(4000) index as a stellar age-sensitive
indicator. Therefore, at the zeroth order, we focus on the subsample of
{\sl quiescent} galaxies in the upper and lower portions of the GV.
The difference in stellar age of the quiescent galaxies serves as a
proxy of the time galaxies spend on the GV. We now resort to stacking
all the quiescent spectra from SDSS, following standard procedures
\citep[see, e.g.][]{Ferreras:13}.
Within a given bin of velocity dispersion, we create one stack in the
upper GV and another one in the lower GV.  The resulting spectra
-- mostly featuring S/N above 100 per \AA\ in the regions of interest -- 
are analysed by measuring a set of age- and metallicity-sensitive spectral
indices: H$\beta$, Mgb and $\langle$Fe$\rangle\equiv$(Fe5270+Fe5335)/2
\citep{Trager:98}; H$\gamma_F$ and H$\delta_F$
\citep{WO:97}; [MgFe]$^\prime$ \citep{TMB:03} and the 4,000\AA\
break strength \citep{Balogh:99}. We include a correction for
nebular emission in the Balmer lines, following the methodology
laid out in appendix B of \cite{FLB:13}. The observed line strengths are
compared with synthetic stellar population models \citep{MIU:12},
producing SSP-equivalent ages (more details will be provided in
a follow-up paper to be published in the main journal).
The top panels of Figure~\ref{fig:SSP_age} show the SSP-equivalent
ages of uGV and lGV galaxies with respect to $\sigma$, and the bottom panels
plot the age difference between these two regions ($\Delta t$), that can be
interpreted as a effective quenching timescale. We emphasize that
these differences concern GV galaxies 
spectroscopically classified as quiescent. The well-known mass-age trend is
readily apparent, with more massive galaxies (higher $\sigma$)
featuring older ages \citep{Gallazzi:05}. It is worth noticing that
both the colour and break strength selection produce very similar
trends. However at the highest values of velocity dispersion 
($\sigma$$>$190\,km\,s$^{-1}$) the dust-corrected colour selection seems
to provide, as absolute SSP ages, older galaxies.

A clearer visualization of this putative GV ``traversing time'' can be
provided by the difference between SSP ages between the uGV and the
lGV.  The data feature a trend of increasing $\Delta$t with velocity
dispersion, implying the GV transit times in more massive galaxies
become longer, from $\Delta$t$\sim$1.5\,Gyr at $\sigma$=100\,\kms\ to
$\sim$3.5\,Gyr at $\sigma$=200\,\kms. These values are roughly in line
with recent estimates of observed and modelled GV
galaxies \citep{Smethurst:15,Wright:18,Rowlands:18}, and our analysis
suggests a clear trend with respect to velocity dispersion.

Intriguingly, the figure suggests a mass scale -- corresponding to the
highest velocity dispersion of the stacks -- above which the age
difference $\Delta$t {\sl decreases}, towards GV transition
times around 1\,Gyr at $\sigma$=250\,\kms. These trends  can 
also be found in M07, where a high fraction of
AGN with high luminosity is suggestive of short quenching
timescales. In line with M07, we also see a slower 
quenching regime at lower velocity dispersion 
($\sigma<$200\,km/s). Such behaviour can be interpreted as 
two different modes of quenching on either side of this mass scale
(if we assume that the transition always proceeds from BC to RS),
or the presence of a substantial amount of rejuvenation in higher mass
galaxies.  Rejuvenation might be due to positive feedback from AGN.
This is expected to play a role in early gas-rich phases of massive
galaxies, where AGN outflows and precessing jets are likely to
overpressurise interstellar gas clouds and induce star formation as in
Cen~A \citep{Crokett:12,McKinley:18,Keel:19}.  The simple, clear-cut
definition of GV galaxies presented in this letter provides strong
constraints on numerical models of galaxy
formation \citep{Gabor:11,Schaye:15,Dubois:16,Springel:18}.

%%%%%%%%%%%%%%%%%%%%%%%%%%%%%%%%%%%%%%%%%%%%%%%%%%%%%%%%%%%%%%%%%%%%%
%%%%%%%%%%%   FIGURE 4
%%%%%%%%%%%%%%%%%%%%%%%%%%%%%%%%%%%%%%%%%%%%%%%%%%%%%%%%%%%%%%%%%%%%%
%%%%%%%%%%%%%%%%%%%%%%%%%%%%%%%%%%%%%%%%%%%%%%%%%%%%%%%%%%%%%%%%%%%%%

%%%%%%%%%%%%%%%%%%%%%%%%%%%%%%%%%%%%%%%%%%%%%%%%
%%%%%%%%%%%%%%%   REFERENCES   %%%%%%%%%%%%%%%%%
%%%%%%%%%%%%%%%%%%%%%%%%%%%%%%%%%%%%%%%%%%%%%%%%
\bibliographystyle{mnras}
\bibliography{MNGVal}

\label{lastpage}

\end{document}